\documentclass{amsart}

\usepackage{cite}
\usepackage{hyperref}
\usepackage{amsmath,amssymb,amsfonts}
\usepackage{algpseudocode}
\usepackage{algorithm}
\usepackage{graphicx}
\usepackage{textcomp}
\usepackage{xcolor}
\usepackage{makecell}

\def\BibTeX{{\rm B\kern-.05em{\sc i\kern-.025em b}\kern-.08em
    T\kern-.1667em\lower.7ex\hbox{E}\kern-.125emX}}
\begin{document}

\newtheorem{theorem}{Theorem}[section]
\newtheorem{lemma}[theorem]{Lemma}
\newtheorem{proposition}[theorem]{Proposition}
\newtheorem{corollary}[theorem]{Corollary}

\newtheorem{definition}[theorem]{Definition}
\newtheorem{example}[theorem]{Example}

\newtheorem{remark}[theorem]{Remark}

\newcommand{\rk}{\operatorname{rk}}

\title{Exploring Commutative Matrix Multiplication Schemes via Flip Graphs}

\author{Isaac Wood}

\address{Institute for Algebra, Johannes Kepler University, Linz, Austria}
\email{isaac.wood@jku.at}

\thanks{Supported by the Austrian FWF grant 10.55776/PAT8258123}

\maketitle

\begin{abstract}
We explore new approaches for finding matrix multiplication algorithms in the commutative setting by adapting the flip graph technique: a method previously shown to be effective for discovering fast algorithms in the non-commutative case. While an earlier attempt to apply flip graphs to commutative algorithms saw limited success, we overcome both theoretical and practical obstacles using two strategies: one inspired by Marakov’s algorithm to multiply 3x3 matrices, in which we construct a commutative tensor and approximate its rank using the standard flip graph; and a second that introduces a fully commutative variant of the flip graph defined via a quotient tensor space. We also present a hybrid method that combines the strengths of both. Across all matrix sizes up to 5x5, these methods recover the best known bounds on the number of multiplications and allow for a comparison of their efficiency and efficacy. Although no new improvements are found, our results demonstrate strong potential for these techniques at larger scales.
\end{abstract}

\section{Introduction}

The computational cost of matrix multiplication is a famous open problem in algebraic complexity theory. In 1969, Strassen  \cite{strassen} discovered that the $n^3$ multiplications needed to multiply matrices isn't optimal, and found an algorithm to multiply $2\times 2$ matrices in only $7$ multiplications, leading to an algorithm for multiplying $n\times n$ matrices in $O(n^{2.805\dots})$ ground field operations. This led to extensive research in similar algorithms, with the most recent optimization finding multiplication being done in $O(n^{2.37133\dots})$ by Alman et al. \cite{alman}.

All these discoveries rely on the algorithms that do not assume commutativity of multiplication, which is crucial for using the algorithm recursively; for instance, Strassen's algorithm computes the product of $2^k \times 2^k$ matrices in $7^k$ field multiplications. However, leveraging commutativity can yield algorithms that use fewer multiplications. Notably, Winograd \cite{winograd} found an algorithm using around $n^3/2$ multiplications that exploited commutativity in 1967, two years earlier than Strassen's algorithm. This commutative algorithm was improved upon by Waksman in 1970 \cite{waksman}, and this remains one of the fastest practical ways to multiply small matrices \cite{drevet}.

Since then, there has been substantial progress in commutative algorithms, we particularly note Marakov's  \cite{marakov} algorithm for multiplying $3\times 3$ matrices in only 22 multiplications, one fewer than Laderman's \cite{laderman} algorithm without commutativity which uses 23 multiplications. Rosowski \cite{rosowski} showed that to multiply an $l\times m$ matrix by a $m\times n$ matrix, you need only $(m(ln+l+n-1)+n-1)/2$ multiplications if $m$ is odd and $n$ is even, and otherwise you need only $m(ln+l+n-1)/2$ multiplications.

For non-commutative algorithms, there have been advancements in searches for the fewest multiplications for specific small sizes. Smirnov \cite{smirnov} used a numerical approach to find improvements in several sizes, Kauers et al. \cite{kauerssat} used SAT-solvers to find many $3\times 3$ multiplication schemes matching Laderman's 23 multiplications, and very recently Google's AlphaEvolve \cite{alphaevolve} found improvements in several sizes as well using its own method, including a $4\times 4$ scheme with 48 multiplications using non-real coefficients: a result now matched over rings with $1/2$ by Dumas et al. \cite{dumas}. The recent method we will be focusing on is the flip graph. This approach involves traversing a connected graph of all Strassen-like algorithms, in the hope of finding one that needs very few multiplications. This approach has yielded significant advancements, as demonstrated in works by Kauers, Moosbauer, Poole, Wood, Arai, Ichikawa and Hukushima \cite{origflip,araietal,jakobnew, me}.

To our knowledge, no computer searches have been successfully applied to find commutative schemes. Our work here uses flip graphs to find commutative algorithms. Previous attempts to apply flip graphs to the commutative setting by Moosbauer \cite{jakobdiss} have been unsuccessful. He suggests there may have been a bug in his implementation, not allowing for rectangular matrix multiplications to work properly. We also tackle this differently from Moosbauer, using three approaches. One in which we can use the commutativity of the multiplications within a generalized flip graph, one in which we use the commutativity before applying a flip graph search in a clever way inspired by Marakov's $3\times 3$ algorithm and then we attempt to combine these approaches to find a fast and accurate method.

We compare the three methods by using them to find bounds on the number of multiplications needed for all matrix sizes up to multiplying two $5\times 5$ matrices. In these sizes, the best known bound is given by a slight improvement to Rosowski's bound we found, noting that $AB=(B^TA^T)^T$. We find that using the standard flip graph search on a modified tensor is about 30 times faster than using the commutative tensor, though this comes at the cost of not matching Rosowski's bound in all matrix sizes, while the commutative flip graph was able to match this bound in every case. The combined approach achieved a favourable balance between efficacy and speed, matching Rosowski's bound in almost all matrix sizes and being around three times faster than the full commutative flip graph approach. Overall, these computational methods were able to match Rosowski's rank in every size of matrix multiplication we looked at and showed great promise for finding improvements on these ranks for bigger sizes, though unfortunately we did not find improvements for any sizes we looked at.

\section{Background}
Let $\mathbb{F}$ be a field, $R$ be an $\mathbb{F}$-algebra and $\mathbf{A}\in R^{l\times m},\mathbf{B} \in R^{m\times n}$. Recall that a Strassen-like algorithm for computing $\mathbf{AB}$ works in two steps: first we perform $r$ multiplications in $R$, then we take some linear combinations of these products to obtain the entries of $\mathbf{AB}$. For example, in Strassen's algorithm, we have $l=m=n=2$: let 
$$\mathbf{A}=\begin{pmatrix}
    A_{11} & A_{12} \\
    A_{21} & A_{22}
\end{pmatrix},\mathbf{B}=\begin{pmatrix}
    B_{11} & B_{12} \\
    B_{21} & B_{22}
\end{pmatrix},\mathbf{AB}=\begin{pmatrix}
    C_{11} & C_{12} \\
    C_{21} & C_{22}
\end{pmatrix}$$
so we get
\begin{align*}
    m_1 &= (A_{11}+A_{22})(B_{11}+B_{22})\\
    m_2 &= (A_{21}+A_{22})B_{11}\\
    m_3 &= A_{11}(B_{12} - B_{22})\\
    m_4 &= (A_{12}+A_{11})B_{22}\\
    m_5 &= A_{22}(B_{21} - B_{11})\\
    m_6 &= (A_{21}-A_{11})(B_{11}+B_{12})\\
    m_7 &= (A_{12}-A_{22})(B_{22}+B_{21})\\
    C_{11}&=m_1-m_4+m_5+m_7\\ C_{12}&=m_3+m_4\\ C_{21}&=m_2+m_5\\ C_{22}&=m_1-m_2+m_3+m_6.
\end{align*}

This allows us to compute $\mathbf{AB}$ using only 7 multiplications.

We often represent these algorithms using the language of tensors, which is formalized as follows:

\begin{definition}
    Let $l,m,n\in\mathbb{N}$. Let $V_{1,l,m}=\mathbb{F}^{l\times m},V_{2,m,n}=\mathbb{F}^{m\times n},V_{3,n,l}=\mathbb{F}^{n\times l}$ be vector spaces over $\mathbb{F}$. Let $\{a_{ij}:  1\leq i \leq l, 1\leq j \leq m\}$ be a basis for $V_{1,l,m}$, $\{b_{ij}:  1\leq i \leq m, 1\leq j \leq n\}$ be a basis for $V_{2,m,n}$ and $\{c_{ij}:  1\leq i \leq n, 1\leq j \leq l\}$ be a basis for $V_{3,n,l}$.
    
    Define the \textit{matrix multiplication tensor} $\mathcal{M}_{l,m,n} \in V_{1,l,m}\otimes V_{2,m,n} \otimes V_{3,n,l}$ to be 
    
    $$\mathcal{M}_{l,m,n} = \sum_{i,j,k=1}^{l,m,n} a_{ij}\otimes b_{jk} \otimes c_{ki}.$$
\end{definition}

\begin{remark}
    This tensor in fact does encode matrix multiplication. Let $A,B,C$ be matrix spaces over $\mathbb{F}$ with dimension $l\times m$, $m \times n$ and $l \times n$. We can now define an isomorphism $\Phi : V_{1,l,m}\otimes V_{2,m,n} \otimes V_{3,n,l} \rightarrow \text{Bil}(A,B;C)$ defined by 
    $$\Phi(a_{i_1j_1}\otimes b_{j_2k_1}\otimes c_{k_2i_2}) = \phi_{i_1,i_2,j_1,j_2,k_1,k_2}$$
    where
    $$\phi_{i_1,i_2,j_1,j_2,k_1,k_2}(\alpha,\beta) = \alpha_{i_1j_1}\beta_{j_2k_1} \Gamma_{i_2k_2}$$
    where $\alpha \in A, \beta \in B, \Gamma_{i_2k_2} \in C$; $\alpha_{ij},\beta_{ij}$ represent the elements in $\alpha$ or $\beta$ respectively in the $(i,j)$-th entry; $\Gamma_{ij}$ is the matrix with all zeroes except a one at the $(i,j)$-th entry.

    Note that the $c_{ij}$ correspond to $\Gamma_{ji}$; this is standard to keep the cyclic nature of the definition of matrix multiplication. We can now write Strassen's algorithm using this tensor notation:
    \begin{align*}
        &(a_{11}+a_{22})\otimes(b_{11}+b_{22})\otimes(c_{11}+c_{22}) + \\
        &(a_{21}+a_{22})\otimes b_{11}\otimes(c_{12}-c_{22}) + \\
        &a_{11}\otimes(b_{12}-b_{22})\otimes(c_{21}+c_{22}) + \\
        &(a_{12}-a_{22})\otimes(b_{21}+b_{22})\otimes c_{11} + \\
        &(a_{12}+a_{11})\otimes b_{22}\otimes(c_{21}-c_{11}) + \\
        &a_{22}\otimes(b_{21}-b_{11})\otimes(c_{12}+c_{11}) + \\
        &(a_{21}-a_{11})\otimes(b_{12}+b_{11})\otimes c_{22}
    \end{align*}
\end{remark}

The following definition will be given for a more general tensor than just matrix multiplication tensors. This generality will be used later.

\begin{definition}
    Given vector spaces $U_1,U_2,U_3$ over $\mathbb{F}$, define a \textit{rank one tensor} to be any tensor in $U_1\otimes U_2 \otimes U_3$ of the form $u_1\otimes u_2 \otimes u_3$ where $u_1\in U_1, u_2\in U_2, u_3 \in U_3$. The \textit{rank} of a tensor $\mathcal{T}\in U_1\otimes U_2 \otimes U_3$ is the smallest $r\in \mathbb{N}$ such that we can write $\mathcal{T} = \sum_{p=0}^r t_p$ where each $t_p$ is a rank one tensor; denote this by $\rk(\mathcal{T})$.
\end{definition}

\begin{remark}
    Investigating the rank of tensors is notoriously difficult. In fact, exact ranks for $\mathcal{M}_{l,m,n}$ is known exactly only if $1\in \{l,m,n\}$, two of $l,m,n$ are 2 or $\{l,m,n\}=\{2,3\}$.
\end{remark}

\begin{definition}
    An \textit{$(l,m,n)$ scheme} for matrix multiplication is a set of $r\in \mathbb{N}$ rank one tensors $\{t_1,\dots,t_r\}$ such that $\mathcal{M}_{l,m,n} = \sum_{p=1}^r t_p$. The rank of such a scheme is its size $r$. We write $T_{l,m,n} = \{a_{ij}\otimes b_{jk}\otimes c_{ki} : 1\leq i \leq l, 1 \leq j \leq m, 1 \leq k \leq n\}$ for the standard $(l,m,n)$ scheme.
\end{definition}

\begin{remark}
    Note that, from the map between the tensors and matrix multiplication algorithms, the rank of an $(l,m,n)$ scheme for matrix multiplication corresponds to the number of multiplications in $R$ needed to compute the matrix multiplication, and hence $\rk(\mathcal{T})$ corresponds to the fewest multiplications in $R$ needed to compute the matrix multiplication using a Strassen-like algorithm.
\end{remark}

We now recall a method to find schemes of low rank from Kauers and Moosbauer \cite{origflip}.

\begin{definition}\label{flip}
    Given an $(l,m,n)$ scheme $S$ of rank $r$, we say the $(l,m,n)$ scheme $S'$ of rank $r$ is a $\textit{flip}$ of $S$ if
    \begin{enumerate}
    \item $$\exists a,b_1,b_2,c_1,c_2\text{ such that }\{a\otimes b_1\otimes c_1,a\otimes b_2\otimes c_2\}\subset S$$
    \item \begin{align*}
S' &= \left( S \backslash \{a\otimes b_1\otimes c_1,a\otimes b_2\otimes c_2\}\right)\\
&\quad \cup \{a\otimes (b_1+ b_2)\otimes c_1,a\otimes b_2\otimes (c_2- c_1)\}
\end{align*}
    \end{enumerate}
    or any permutation between $a,b,c$.
\end{definition}

\begin{definition}\label{reducible}
    We say an $(l,m,n)$ scheme $S=\{a_i\otimes b_i\otimes c_i|i\in\{1,\dots,r\}\}$ of rank $r$ over $\mathbb{F}$ is \textit{reducible} if there is some subset $I\subset \{1,\dots,r \}$ with $a_i=\alpha_i a$ for all $i\in I$ and $\sum_{i\in I\backslash\{t\}} \beta_i b_i=b_t$, where $a\in \mathbb{F}^{l\times m}$, $t\in I$ and $\alpha_i,\beta_i\in\mathbb{F}$ for all $i\in I$.
    We call the $(l,m,n)$ scheme $S'$ a \textit{reduction} of $S$ where
    \begin{align*}
S'&=\{a_j\otimes b_j\otimes c_j:j\in\{1,\dots,r\}\backslash I\}\\
&\quad \cup \{ a_i\otimes b_i \otimes ( c_i + \alpha_i\beta_i c_t ) : i\in I\backslash\{t\}\}
\end{align*}
\end{definition}

\begin{definition}
    Let $V_{l,m,n}$ be the set of all $(l,m,n)$ schemes and $E_{l,m,n}$ be the set of  pairs of schemes $\{S_1,S_2\}$ where $S_1$ is a flip of $S_2$, $S_1$ is a reduction of $S_2$, or $S_2$ is a \textit{reduction} of $S_1$. We call the graph $(V_{l,m,n},E_{l,m,n})$ the \textit{$(l,m,n)$ flip graph}.
\end{definition}

\begin{theorem}
     \cite{origflip} The $(l,m,n)$ flip graph is connected.
\end{theorem}

Kauers and Moosbauer traverse this graph randomly from the standard scheme $T_{l,m,n}$ to try to find schemes of low rank. In doing so, they found improvements to a few sizes of schemes, and were able to match the best known in all other sizes up to $(5,5,5)$.

\begin{remark}
    In implementations, it seems to be more efficient to not check for reductions and instead wait for the scheme to be flipped into a ``trivial reduction'' (i.e one of the elements of the scheme is $0$). This is because checking for a non-trivial reduction is costly, and when doing a random flip you are mostly unlikely to destroy a possible reduction. The time saved by doing a few less flips is much less than that saved by not checking for a non-trivial reduction.
\end{remark}
\begin{remark}
    Usually when these algorithms are implemented, we perform the flip graph over $\mathbb{Z}_2$ largely for simplicity and efficiency, but if we want schemes over $\mathbb{Z}$ we can try Hensel lifting the schemes we find.
\end{remark}
The original search method from Kauers and Moosbauer was improved upon by Arai et al \cite{araietal} to include two new ideas: first they include a new \textit{plus} operation, which increases the rank of a scheme, avoiding the search finding a non-optimal local minimum; second they don't start the search from $T_{l,m,n}$ but instead start from a scheme constructed from a fast scheme of a smaller size. These improvements to the search drastically decreased computation time, allowing Arai et al. to reproduce Kauers and Moosbauer's results (and even find a couple of improvements) in under an hour using a laptop.

\begin{definition}\label{plus}
    Given $(l,m,n)$ schemes $S,S'$ with $S'$ with rank 1 more than that of $S$, we call $S'$ a \textit{plus} of $S$ if $\exists a_1\otimes b_1 \otimes c_1,a_2\otimes b_2 \otimes c_2 \in S$ such that 
    \begin{align*}
S'&=S\backslash\{ a_1\otimes b_1 \otimes c_1,a_2\otimes b_2 \otimes c_2 \}\\
&\quad\quad \cup \{ a_1\otimes (b_1+b_2)\otimes c_1, (a_2-a_1)\otimes b_2 \otimes c_2,\\
&\quad\quad\quad\quad a_1 \otimes b_2 \otimes (c_2 - c_1) \}
\end{align*}
    or
    \begin{align*}
S'&=S\backslash\{ a_1\otimes b_1 \otimes c_1,a_2\otimes b_2 \otimes c_2 \}\\
&\quad\quad \cup \{ a_1\otimes b_1\otimes (c_1+c_2), a_2\otimes (b_2-b_1)\otimes c_2,\\
&\quad\quad\quad\quad(a_2 - a_1)\otimes b_1 \otimes c_2 \}
\end{align*}
    or 
    \begin{align*}
S'&=S\backslash\{ a_1\otimes b_1 \otimes c_1,a_2\otimes b_2 \otimes c_2 \}\\
&\quad\quad \cup \{ (a_1+a_2)\otimes b_1\otimes c_1, a_2\otimes b_2\otimes (c_2-c_1),\\
&\quad\quad\quad\quad a_2 \otimes (b_2 - b_1)\otimes c_1\}
\end{align*}
\end{definition}

A plus of a scheme $S$ is simply a scheme $S'$ which is similar to $S$ but has one higher rank.

\begin{definition}
    Let $V_{l,m,n}$ be the set of $(l,m,n)$ schemes and $E_{l,m,n}$ be the set of ordered pairs of schemes $(S_1,S_2)$ where $S_2$ is a flip of $S_1$, $S_2$ is a plus of $S_1$ or $S_2$ is a reduction of $S_1$. We call the directed graph $(V_{l,m,n},E_{l,m,n})$ the \textit{adaptive $(l,m,n)$ flip graph}.
\end{definition}

\begin{theorem}
     \cite{araietal}
    The adaptive $(l,m,n)$ flip graph over $\mathbb{F}_2$ is connected.
\end{theorem}

An algorithm using this new flip graph is given below in Algorithm \ref{adaptivenc}.

\begin{algorithm}
\caption{Adaptive Flip Graph Search}\label{adaptivenc}
\begin{algorithmic}[1]
\State \textbf{input:} $S$ an $(l,m,n)$ scheme
\For {some number of iterations}
    \If{$S$ is reducible}
        \State $S \gets$ reduction of $S$
    \EndIf
    \If{we cannot do a flip from $S$}
        \State {$S\gets$ a plus of $S$}
    \EndIf
    \State $S \gets$ flip of $S$
    \State occasionally $S\gets$ a plus of $S$
\EndFor
\State \Return the best scheme seen in the search
\end{algorithmic}
\end{algorithm}

Arai et al note that we can consider $V_{1,l-1,m}\otimes V_{2,m,n} \otimes V_{3,n,l-1} \subset V_{1,l,m} \otimes V_{2,m,n} \otimes V_{3,n,l}$, hence given an $(l-1,m,n)$ scheme $\mathcal{S}$ of low rank we can see that

$$\mathcal{S}\cup \{ a_{lj} \otimes b_{jk} \otimes c_{kl} : 1\leq j \leq m, 1 \leq k \leq n \}$$

is an $(l,m,n)$ scheme of non-trivial rank too. This technique allows Arai et al. to construct effective starting points for the search.

For focusing on commutative algorithms, we will compare our results to the following theorem from Rosowski:

\begin{theorem}
     \cite{rosowski} Given $l,m,n\in\mathbb{N}$, if $m$ is odd and $n$ is even, then there exists an algorithm to multiply an $l\times m$ matrix by an $m\times n$ matrix using $(m(ln+l+n-1)+l-1)/2$ multiplications, otherwise there exists an algorithm to multiply an $l\times m$ matrix by an $m\times n$ matrix using $m(ln+l+n-1)/2$ multiplications.
\end{theorem}

We note that this can be improved slightly, as since $(AB)^T=B^TA^T$ we see that the existence of an algorithm to multiply an $l\times m$ matrix by an $m \times n$ matrix in $r$ multiplications gives us the existence of an algorithm to multiply an $n\times m$ matrix by an $m \times l$ matrix in $r$ multiplications.

\begin{corollary}
    There exists an algorithm to multiply an $l\times m$ matrix by an $m \times n$ matrix in $r$ multiplications, where
    $$r= \begin{cases}
        \begin{split}
        (m(ln+l+n-1)&+\\\min(l,n)-1)/2
        \end{split}
        & \text{if $m$ odd, $l,n$ both even} \\
        m(ln+l+n-1)/2 & \text{else}
    \end{cases}$$
\end{corollary}

These are the best known bounds for small $l,m,n$, though note we can even find better commutative schemes for larger $l,m,n$ e.g with Strassen's algorithm.

\section{Flip Graph Searches for Commutative Schemes}

In this section we present new methods for finding commutative schemes of low rank based on the flip graph. We aim to use flip graphs methods, but note that we cannot directly apply these into the commutative setting, since a standard flip graph search can only ever yield non-commutative schemes. As such we present two methods, and consider combining the two methods together for faster implementations.

\subsection{Extending Marakov's idea}

In this first approach, we use the commutativity of multiplication before applying the flip graph so that we can keep using the standard non-commutative flip graph search. We consider Marakov's 1986 paper \cite{marakov}: he presents an algorithm for multiplying $3\times3$ matrices using only 22 multiplications. The products he used were found by partitioning the elements of the matrices into two disjoint sets, and then each of his multiplications are a linear combination of elements of one set with a linear combination of elements of the other set. More specifically, we take

$$\mathbf{A}=\begin{bmatrix}
\textcolor{red}{A_{11}^{[1]}} & \textcolor{blue}{A_{12}^{[2]}} & \textcolor{blue}{A_{13}^{[2]}} \\
\textcolor{red}{A_{21}^{[1]}} & \textcolor{blue}{A_{22}^{[2]}} & \textcolor{blue}{A_{23}^{[2]}} \\
\textcolor{red}{A_{31}^{[1]}} & \textcolor{blue}{A_{32}^{[2]}} & \textcolor{blue}{A_{33}^{[2]}}
\end{bmatrix}, \quad \mathbf{B}=\begin{bmatrix}
\textcolor{blue}{B_{11}^{[2]}} & \textcolor{blue}{B_{12}^{[2]}} & \textcolor{blue}{B_{13}^{[2]}} \\
\textcolor{red}{B_{21}^{[1]}} & \textcolor{red}{B_{22}^{[1]}} & \textcolor{red}{B_{23}^{[1]}} \\
\textcolor{red}{B_{31}^{[1]}} & \textcolor{red}{B_{32}^{[1]}} & \textcolor{red}{B_{33}^{[1]}}
\end{bmatrix}$$

where $^{\textcolor{red}{[1]}}$ and $ ^{\textcolor{blue}{[2]}}$ are the two sets, and every multiplication is between a linear combination of these. As in the non-commutative case, we can express this in the language of tensors: we take vector spaces $\textcolor{red}{V_1'}=\mathbb{F}^9,\textcolor{blue}{V_2'}=\mathbb{F}^9,V_3'=\mathbb{F}^9$ with $\{ a_{11},a_{21},a_{31},b_{21},b_{22},b_{23},b_{31},b_{32},b_{33} \}$ a basis for $V_1'$, $\{ a_{12},a_{13},a_{22},a_{23},a_{32},a_{33},b_{11},b_{12},b_{13} \}$ a basis for $V_2'$ and $\{ c_{11},c_{12},c_{13},c_{21},c_{22},c_{23},c_{31},c_{32},c_{33} \}$ a basis for $V_3'$. Marakov showed that 
$$\rk\left( \sum_{i,k=1}^3 \left( \sum_{j=2}^3 b_{jk} \otimes a_{ij} \otimes c_{ki}\right)+ a_{i1} \otimes b_{1k}\otimes c_{ki} \right)\leq 22.$$

After some experimenting, we found the following tensor that represents a commutative matrix multiplication scheme especially fruitful:

\begin{definition}
    Let $l,m,n\in\mathbb{N}$ and let $V_{1,l,m}'= \mathbb{F}^{l\times\lfloor m/2\rfloor},V_{2,m,n}=\mathbb{F}^{\lfloor m/2 \rfloor \times n}, V_{3,n,l}' =\mathbb{F}^{n\times l}$ be vector spaces generated by \begin{align*}
&\{ a_{ij} : 1\leq i \leq l, 1\leq j \leq m, j\text{ even} \}\\
&\quad\cup\{ b_{ij} : 1\leq i \leq m, 1 \leq j \leq n, i\text{ even} \},
\end{align*} \begin{align*}
&\{ a_{ij} : 1\leq i \leq l, 1\leq j \leq m, j\text{ odd} \}\\
&\quad\cup\{ b_{ij} : 1\leq i \leq m, 1 \leq j \leq n, i\text{ odd} \}
\end{align*} and $$\{ c_{ij} : 1\leq i \leq n, 1 \leq j \leq l \}$$ respectively. We define the \textit{$(l,m,n)$ Marakov-like matrix multiplication tensor} to be
    \begin{align*}
        \mathcal{V}_{l,m,n}&=\sum_{i,k=1}^{l,n} \left( \sum_{\substack{j=1 \\ j\text{ odd}}}^m a_{ij}\otimes b_{jk} \otimes c_{ki} \right. \\
        &\quad\quad\quad\quad\quad \left. {}+ \sum_{\substack{j=2 \\ j\text{ even}}}^m b_{jk}\otimes a_{ij} \otimes c_{ki}  \right) \\
        &\quad\quad\quad\quad\quad\quad\quad \in V_{1,l,m}'\otimes V_{2,m,n}'\otimes V_{3,n,l}'
    \end{align*}
    We define an \textit{$(l,m,n)$ Marakov-like scheme} to be a set of $r$ rank one tensors whose sum is $\mathcal{V}_{l,m,n}$; the rank of a Marakov-like scheme is the size $r$ of the set.
\end{definition}

We can now directly apply Algorithm \ref{adaptivenc} to find a low rank for Marakov-like schemes, and this will give us a (hopefully good) upper bound on the commutative rank.

As before, we can also take an $(l-1,m,n)$ Marakov-like scheme and then ``extend'' it to an $(l,m,n)$ Marakov-like scheme to find better starting points than the standard algorithm.

\subsection{Commutative flip graph}

We will define what we mean by a commutative tensor and how we will use them to represent commutative matrix multiplications. Perhaps the following definition may seem a little formal and far removed from intuition, but the quotient space simply captures the idea that $a\otimes b\otimes c$ and $b\otimes a\otimes c$ should be the same in the commutative setting.

\begin{definition}
    Let $l,m,n\in\mathbb{N}$ and let $U_1 = \mathbb{F}^{l\times m + m\times n}$ be an $\mathbb{F}$ vector space with basis $\{ a_{ij} : 1\leq i\leq l,1\leq j\leq m \} \cup \{ b_{ij} : 1\leq i \leq m, 1\leq j \leq n \}$ and let $U_2 = \mathbb{F}^{n\times l}$ be an $\mathbb{F}$ vector space with basis $\{ c_{ij} : 1\leq i \leq n, 1\leq j \leq l \}$. Let $S=\text{span}\{ u_{11}\otimes u_{12} \otimes u_2 - u_{12}\otimes u_{11} \otimes u_2 : u_{11},u_{12}\in U_1, u_2\in U_2 \} $, and now let $U=(U_1\otimes U_1 \otimes U_2) / S$ be the quotient space.
    
    We define the \textit{matrix multiplication commutative tensor} to be 
    $$\mathcal{M}_{l,m,n}' = \sum_{i,j,k=1}^{l,m,n}a_{ij}\otimes b_{jk}\otimes c_{ki} + S\in U.$$
    Similarly, we now define a \textit{rank one commutative tensor} to be any tensor of the form $\alpha \otimes \beta \otimes\gamma + S\in U$, and we again define the rank of $\mathcal{M}_{l,m,n}'$, written $\rk'(\mathcal{M}_{l,m,n})$, to be the smallest $r$ such that $\mathcal{M}_{l,m,n}'=\sum_{i=1}^r t_i$ with each $t_i$ a rank one commutative tensor. We also define an \textit{$(l,m,n)$ commutative scheme} to be a set of $r$ rank one commutative tensors whose sum is $\mathcal{M}_{l,m,n}'$, and the \textit{rank} of an $(l,m,n)$ commutative scheme is the size $r$ of this set.
\end{definition}

We aim to extend the flip graph to commutative schemes, such that we can extend Algorithm \ref{adaptivenc} to find commutative schemes.

\begin{definition}
    The definitions of \textit{flip}, \textit{reducible}, and \textit{plus} for commutative schemes are direct analogues of Definitions \ref{flip}, \ref{reducible}, and \ref{plus}, applied to elements in the quotient space of commutative tensors.
    Let $V'_{l,m,n}$ be the set of $(l,m,n)$ commutative schemes and $E'_{l,m,n}$ be the set of  ordered pairs of commutative schemes $\{S_1,S_2\}$ where $S_2$ is a flip of $S_1$, $S_2$ is a plus of $S_1$ or $S_2$ is a reduction of $S_1$. We call the directed graph $(V'_{l,m,n},E'_{l,m,n})$ the \textit{commutative adaptive $(l,m,n)$ flip graph}.
\end{definition}

\begin{remark}
    Note that although the definitions here are almost identical to the non-commutative definitions, implementations will be rather different since in the commutative case there are more ways to flip, e.g $a_1\otimes b_1\otimes c+a_2\otimes b_2\otimes c$ can flip in four distinct ways instead of the two ways in the non-commutative case, namely:
    \begin{enumerate}
        \item $a_1 \otimes (b_1 \pm b_2)\otimes c+(a_2 \mp a_1)\otimes b_2 \otimes c$
        \item $(a_1 \pm a_2) \otimes b_1 \otimes c + a_2 \otimes (b_1 \mp b_2) \otimes c$
        \item $a_1 \otimes (b_1 \pm a_2)\otimes c+(a_2 \mp b_1)\otimes b_2 \otimes c$
        \item $(a_1 \pm b_2) \otimes b_1 \otimes c + (a_2 \mp b_1) \otimes b_2 \otimes c$.
    \end{enumerate}
\end{remark}

\begin{theorem}
    The commutative adaptive $(l,m,n)$ flip graph over $\mathbb{F}_2$ is connected.
\end{theorem}

The proof of this theorem exactly follows the corresponding proof from the non-commutative case.

We can hence extend Algorithm \ref{adaptivenc} to the commutative case.

\begin{remark}
When implementing this algorithm, we increased the frequency with which we perform the plus operation. This is because in the commutative flip graph, it seems reductions are a bit harder to find, perhaps due to the increased choices with which we can flip, and so we need to do a plus less often to avoid the case where we increase the rank more than we decrease it.
\end{remark}

\subsection{Combination}

One major drawback of the commutative flip graph search is that it is much slower than the standard flip graph search, not least because the implementation for the standard flip graph search has been optimised much more than the commutative flip graph search. This can be countered, however, by using a combination of the two above methods. Taking a Marakov-like scheme of low rank, and using it as a starting point in the commutative flip graph search.

\section{Table of Results}

Here are the resulting best ranks of schemes we found for a commutative $(l,m,n)$ scheme over $\mathbb{F}_2$, along with the rank Rosowski finds, using the three methods listed above. In bold, we can see any places in which the method differs from the best known. It should be noted that none of these methods found an improvement from Rosowski's methods in these sizes in our runs.

\begin{center}
\begin{tabular}{c|c|c|c|c|c|c}
$l$ & $m$ & $n$ & Ros. & M & C & M-C\\
\hline
2 & 2 & 2 & 7 & 7 & 7 & 7 \\
2 & 2 & 3 & 10 & 10 & 10 & 10\\
2 & 2 & 4 & 13 & 13 & 13 & 13\\
2 & 2 & 5 & 16 & 16 & 16 & 16\\
2 & 3 & 2 & 11 & 11 & 11 & 11\\
2 & 3 & 3 & 15 & 15 & 15 & 15\\
2 & 3 & 4 & 20 & 20 & 20 & 20\\
2 & 3 & 5 & 24 & \textbf{25} & 24 & 24\\
2 & 4 & 2 & 14 & 14 & 14 & 14\\
2 & 4 & 3 & 20 & 20 & 20 & 20\\
2 & 4 & 4 & 26 & 26 & 26 & 26\\
2 & 4 & 5 & 32 & 32 & 32 & 32\\
2 & 5 & 2 & 18 & 18 & 18 & 18\\
2 & 5 & 3 & 25 & 25 & 25 & 25\\
2 & 5 & 4 & 33 & 33 & 33 & 33\\
2 & 5 & 5 & 40 & 40 & 40 & 40\\
3 & 2 & 3 & 14 & 14 & 14 & 14\\
3 & 2 & 4 & 18 & 18 & 18 & 18\\
3 & 2 & 5 & 22 & 22 & 22 & 22\\
3 & 3 & 3 & 21 & \textbf{22} & 21 & 21\\
3 & 3 & 4 & 27 & \textbf{28} & 27 & \textbf{28}\\
3 & 3 & 5 & 33 & \textbf{35} & 33 & 33\\
3 & 4 & 3 & 28 & 28 & 28 & 28\\
3 & 4 & 4 & 36 & 36 & 36 & 36\\
3 & 4 & 5 & 44 & 44 & 44 & 44\\
3 & 5 & 3 & 35 & \textbf{36} & 35 & 35\\
3 & 5 & 4 & 45 & \textbf{47} & 45 & \textbf{46}\\
3 & 5 & 5 & 55 & \textbf{58} & 55 & 55\\
4 & 2 & 4 & 23 & 23 & 23 & 23\\
4 & 2 & 5 & 28 & 28 & 28 & 28\\
4 & 3 & 4 & 36 & \textbf{37} & 36 & 36\\
4 & 3 & 5 & 42 & \textbf{45} & 42 & 42\\
4 & 4 & 4 & 46 & 46 & 46 & 46\\
4 & 4 & 5 & 56 & 56 & 56 & 56\\
4 & 5 & 4 & 59 & \textbf{60} & 59 & 59\\
4 & 5 & 5 & 70 & \textbf{73} & 70 & 70\\
5 & 2 & 5 & 34 & 34 & 34 & 34\\
5 & 3 & 5 & 51 & \textbf{55} & 51 & 51\\
5 & 4 & 5 & 68 & 68 & 68 & 68\\
5 & 5 & 5 & 85 & \textbf{89} & 85 & 85\\
\end{tabular}
\end{center}
\vspace{-0.5em} % Adjust spacing if needed
\footnotesize % Use a smaller font size for the key
\begin{itemize}
    \item \textbf{Ros.}: Rosowski's bound
    \item \textbf{M}: Marakov-like tensor with standard flip graph search
    \item \textbf{C}: Commutative flip graph search
    \item \textbf{M-C}: Commutative flip graph search starting from the schemes found in \textbf{M}
\end{itemize}
\normalsize % Revert to normal font size after the key
\bigskip
\bigskip

Using the generalization of Marakov's method is by far the fastest, and this column of the table took 2 hours and 38 minutes to compute using 30 processors, of course this comes at the expense of optimality as it frequently doesn't find the fastest known approach. The second column, though it never failed to match the best known rank, took 50 hours to compute, with tweaking parameters suggesting that this is a typical duration for this method (to achieve results this good at least). This is why we feel the combination worked exceptionally well, given it took around 15 hours to compute and most often matched the best known rank. While the combined approach did not consistently match Rosowski's bounds in all initial runs (i.e the $(3,3,4),(3,5,4)$ entries), extended computation time for these specific cases within the combined method allowed us to eventually match Rosowski's rank. This demonstrates the potential of the combined method to achieve optimal bounds while still offering significant computational speedup compared to a full commutative flip graph search from scratch. As such, this combination method shows excellent potential for finding new bounds.

Using this combination method, we ran a large computation for all sizes up to $(7,7,7)$ to see if we could find any improvements over Rosowski's bounds. Although we found no improvements, we still matched Rosowski's bound in all sizes except some of the form $(a,7,b)$.

%\bibliographystyle{plainurl}
%\bibliography{references}

\begin{thebibliography}{10}

\bibitem{alman}
Josh Alman, Ran Duan, Virginia~Vassilevska Williams, Yinzhan Xu, Zixuan Xu, and
  Renfei Zhou.
\newblock {\em More Asymmetry Yields Faster Matrix Multiplication}, pages
  2005--2039.
\newblock URL: \url{https://epubs.siam.org/doi/abs/10.1137/1.9781611978322.63},
  \href
  {https://arxiv.org/abs/https://epubs.siam.org/doi/pdf/10.1137/1.9781611978322.63}
  {\path{arXiv:https://epubs.siam.org/doi/pdf/10.1137/1.9781611978322.63}},
  \href {https://doi.org/10.1137/1.9781611978322.63}
  {\path{doi:10.1137/1.9781611978322.63}}.

\bibitem{araietal}
Yamato Arai, Yuma Ichikawa, and Koji Hukushima.
\newblock Adaptive flip graph algorithm for matrix multiplication.
\newblock In {\em Proceedings of the 2024 International Symposium on Symbolic
  and Algebraic Computation}, ISSAC '24, page 292–298, New York, NY, USA,
  2024. Association for Computing Machinery.
\newblock \href {https://doi.org/10.1145/3666000.3669701}
  {\path{doi:10.1145/3666000.3669701}}.

\bibitem{dumas}
Jean-Guillaume Dumas, Clément Pernet, and Alexandre Sedoglavic.
\newblock A non-commutative algorithm for multiplying 4x4 matrices using 48
  non-complex multiplications, 2025.
\newblock URL: \url{https://arxiv.org/abs/2506.13242}, \href
  {https://arxiv.org/abs/2506.13242} {\path{arXiv:2506.13242}}.

\bibitem{kauerssat}
Marijn J.~H. Heule, Manuel Kauers, and Martina Seidl.
\newblock Local search for fast matrix multiplication.
\newblock In Mikol{\'a}{\v{s}} Janota and In{\^e}s Lynce, editors, {\em Theory
  and Applications of Satisfiability Testing -- SAT 2019}, pages 155--163,
  Cham, 2019. Springer International Publishing.

\bibitem{origflip}
Manuel Kauers and Jakob Moosbauer.
\newblock Flip graphs for matrix multiplication.
\newblock In {\em Proceedings of the 2023 International Symposium on Symbolic
  and Algebraic Computation}, ISSAC '23, page 381–388, New York, NY, USA,
  2023. Association for Computing Machinery.
\newblock \href {https://doi.org/10.1145/3597066.3597120}
  {\path{doi:10.1145/3597066.3597120}}.

\bibitem{me}
Manuel Kauers and Isaac Wood.
\newblock Consequences of the moosbauer-poole algorithms.
\newblock {\em arXiv preprint arXiv:2505.05896}, 2025.

\bibitem{laderman}
John~D. Laderman.
\newblock A noncommutative algorithm for multiplying 3×3 matrices using 23
  multiplications.
\newblock {\em Bulletin of the American Mathematical Society}, 82:126--128,
  1976.
\newblock \href {https://doi.org/10.1090/S0002-9904-1976-13988-2}
  {\path{doi:10.1090/S0002-9904-1976-13988-2}}.

\bibitem{marakov}
O.M. Makarov.
\newblock An algorithm for multiplying 3×3 matrices.
\newblock {\em USSR Computational Mathematics and Mathematical Physics},
  26(1):179--180, 1986.
\newblock URL:
  \url{https://www.sciencedirect.com/science/article/pii/004155538690203X},
  \href {https://doi.org/https://doi.org/10.1016/0041-5553(86)90203-X}
  {\path{doi:https://doi.org/10.1016/0041-5553(86)90203-X}}.

\bibitem{jakobdiss}
Jakob Moosbauer.
\newblock {\em Search Techniques for Matrix Algorithms}.
\newblock PhD thesis, Johannes Kepler University, 2023.
\newblock URL: \url{https://epub.jku.at/obvulihs/content/titleinfo/9217131}.

\bibitem{jakobnew}
Jakob Moosbauer and Michael Poole.
\newblock Flip graphs with symmetry and new matrix multiplication schemes,
  2025.
\newblock URL: \url{https://arxiv.org/abs/2502.04514}, \href
  {https://arxiv.org/abs/2502.04514} {\path{arXiv:2502.04514}}.

\bibitem{alphaevolve}
Alexander Novikov, Ngân Vũ, Marvin Eisenberger, Emilien Dupont, Po-Sen Huang,
  Adam~Zsolt Wagner, Sergey Shirobokov, Borislav Kozlovskii, Francisco J.~R.
  Ruiz, Abbas Mehrabian, M.~Pawan Kumar, Abigail See, Swarat Chaudhuri, George
  Holland, Alex Davies, Sebastian Nowozin, Pushmeet Kohli, and Matej Balog.
\newblock Alphaevolve: A coding agent for scientific and algorithmic discovery,
  2025.
\newblock URL: \url{https://arxiv.org/abs/2506.13131}, \href
  {https://arxiv.org/abs/2506.13131} {\path{arXiv:2506.13131}}.

\bibitem{rosowski}
Andreas Rosowski.
\newblock Fast commutative matrix algorithms.
\newblock {\em Journal of Symbolic Computation}, 114:302--321, 2023.
\newblock URL:
  \url{https://www.sciencedirect.com/science/article/pii/S0747717122000499},
  \href {https://doi.org/https://doi.org/10.1016/j.jsc.2022.05.002}
  {\path{doi:https://doi.org/10.1016/j.jsc.2022.05.002}}.

\bibitem{smirnov}
A.~V. Smirnov.
\newblock The bilinear complexity and practical algorithms for matrix
  multiplication.
\newblock {\em Computational Mathematics and Mathematical Physics},
  53(12):1781--1795, 2013.
\newblock \href {https://doi.org/10.1134/S0965542513120129}
  {\path{doi:10.1134/S0965542513120129}}.

\bibitem{strassen}
Volker Strassen.
\newblock Gaussian elimination is not optimal.
\newblock {\em Numerische Mathematik}, 13(4):354--356, 1969.
\newblock \href {https://doi.org/10.1007/BF02165411}
  {\path{doi:10.1007/BF02165411}}.

\bibitem{waksman}
A.~Waksman.
\newblock On winograd's algorithm for inner products.
\newblock {\em IEEE Transactions on Computers}, C-19(4):360--361, 1970.
\newblock \href {https://doi.org/10.1109/T-C.1970.222926}
  {\path{doi:10.1109/T-C.1970.222926}}.

\bibitem{winograd}
S.~Winograd.
\newblock A new algorithm for inner product.
\newblock {\em IEEE Transactions on Computers}, C-17(7):693--694, 1968.
\newblock \href {https://doi.org/10.1109/TC.1968.227420}
  {\path{doi:10.1109/TC.1968.227420}}.

\bibitem{drevet}
Charles Éric Drevet, Md. {Nazrul Islam}, and Éric Schost.
\newblock Optimization techniques for small matrix multiplication.
\newblock {\em Theoretical Computer Science}, 412(22):2219--2236, 2011.
\newblock URL:
  \url{https://www.sciencedirect.com/science/article/pii/S0304397510007036},
  \href {https://doi.org/https://doi.org/10.1016/j.tcs.2010.12.012}
  {\path{doi:https://doi.org/10.1016/j.tcs.2010.12.012}}.

\end{thebibliography}

\end{document}